\RequirePackage{fix-cm}
\documentclass[twocolumn]{svjour3}          % twocolumn
\smartqed  % flush right qed marks, e.g. at end of proof
\usepackage{graphicx}
\usepackage{mathptmx}      % use Times fonts if available on your TeX system
\usepackage[italian,english]{babel}
\usepackage{subfig}
\usepackage{amsmath}
\usepackage{amssymb}
\usepackage{mathtools}
\usepackage{color}
\usepackage{times}
\usepackage{braket}
\usepackage{natbib}
\newcommand\blfootnote[1]{%
  \begingroup
  \renewcommand\thefootnote{}\footnote{#1}%
  \addtocounter{footnote}{-1}%
  \endgroup
}
\journalname{Rendiconti Lincei}

\DeclarePairedDelimiter{\abs}{\lvert}{\rvert}

\begin{document}

\title{Coherent perfect absorption in photonic structures%\thanks{Grants or other notes
%about the article that should go on the front page should be
%placed here. General acknowledgments should be placed at the end of the article.}
}
%\subtitle{Do you have a subtitle?\\ If so, write it here}

%\titlerunning{Short form of title}        % if too long for running head

\author{Lorenzo Baldacci         \and
        Simone Zanotto	\and
	Alessandro Tredicucci %etc.
}

%\authorrunning{Short form of author list} % if too long for running head

\institute{L. Baldacci \at
              NEST, CNR Istituto Nanoscienze and Scuola Normale Superiore, Piazza San Silvestro 12, 56127 Pisa (Italy) \\
	   Institute of Life Sciences, Scuola Superiore Sant'Anna, Piazza Martiri della Libert\`a 33, 56127 Pisa (Italy) \\
              %Tel.: +123-45-678910\\
              %Fax: +123-45-678910\\
              \email{l.baldacci@sssup.it}           %  \\
%             \emph{Present address:} of F. Author  %  if needed
           \and
           S. Zanotto \at
              Dipartimento di Elettronica, Informazione e Bioingengeria, Politecnico di Milano, Via Colombo 81 20133 Milano (Italy)
	\and
	A. Tredicucci \at
	   NEST, CNR Istituto Nanoscienze and Dipartimento di Fisica, Universit\`a di Pisa, Largo Pontecorvo 3, 56127 Pisa (Italy)
}

\date{Received: date / Accepted: date}
% The correct dates will be entered by the editor

\maketitle

\blfootnote{This contribution is the written, peer-reviewed version of a paper presented at one of the two conferences “From Life to Life: Through New Materials and Plasmonics” - Accademia Nazionale dei Lincei in Rome on June 23, 2014, and “NanoPlasm 2014: New Frontiers in Plasmonics and NanoOptics” - Cetraro (CS) on June 16-20, 2014.}

\begin{abstract}
The ability to drive a system with an external input is a fundamental aspect of light-matter interaction. The coherent perfect absorption (CPA) phenomenon extends to the general multibeam interference phenomenology the well known critical coupling concepts [\cite{Haus,Yariv}]. The latter detail the conditions under which the energy of the input field is fed in full to the absorbing element. In a multi-port system, the relative phase of the incoming fields can yield the ultimate control of the absorption, from CPA to complete transparency (coherent perfect transparency, CPT), and also beyond, in amplifying regimes, to laser threshold control [\cite{LonghiPRA2010, SunPRL14}]. This interferometric control of absorption can be employed to reach perfect energy feeding into nanoscale systems such as plasmonic nanoparticles [\cite{noh2012CPA}], and multi-port interference can be used to enhance the absorption when they are embedded in a strongly scattering system [\cite{chong2011multiportCPA}], with potential applications to nanoscale sensing.
Here we review the two-port CPA in reference to photonic structures which can resonantly couple to the external fields. A revised two-port theory of CPA is illustrated, which relies on the Scattering Matrix formalism and is valid for all linear two-port systems with reciprocity. Through a semiclassical approach, treating two-port critical coupling conditions in a non-perturbative regime, it is demonstrated that the strong coupling regime and the critical coupling condition can indeed coexist; in this situation, termed strong critical coupling [\cite{ZanottoSperoNatPhys}], all the incoming energy is converted into polaritons. 
Experimental results are presented, which clearly display the elliptical trace of absorption as function of input unbalance in a thin metallo-dielectric metamaterial, and verify polaritonic CPA in an intersubband-polariton photonic-crystal membrane resonator. Concluding remarks discuss the future perspectives of CPA with photonic structures.
% \PACS{PACS code1 \and PACS code2 \and more}
% \subclass{MSC code1 \and MSC code2 \and more}
\end{abstract}

\section{Introduction}
\label{intro}

In opto-electronic devices, energy dissipation can be highly
undesirable or very much needed, depending on the foreseen
application. Photon detectors and solar cells are the prototypical
semiconductor components belonging to the second category: the
incoming electromagnetic energy must be mostly transferred and
dissipated within the active region for optimized operation [\cite{YuPRL2012}]. The
same condition holds for most microwave and radio systems:
an antenna must efficiently deliver/extract the signal to/from the
receiving/emitting circuit. The crucial concept here is the critical
coupling between input channel and load [\cite{Yariv}]. It is nothing else than the
so-called impedance matching condition, which, once expressed in
terms of losses, indeed states that radiative and material losses must
be equal [\cite{Haus}]. On one hand, the critical coupling provides a common
framework to the efforts aiming at developing perfect absorbers
and optimal thermal emitters [\cite{GhebrebrhanPRA2011}]. On the other hand, it is noticeably
similar to the lasing condition, on exchanging losses with gain.
Indeed, the link between perfect absorbers and lasers has been
recently pointed out in terms of time reversal within the coherent
perfect absorption (CPA) framework [\cite{ChongPRL2010}].
CPA consists in the complete dissipation of the input optical
power, occurring when the amplitude and phase of one input beam have a well-defined relation
with the amplitudes and phases of the other beam(s), provided that the determinant of the system's
scattering matrix is vanishing.
The simplest CPA device is a symmetric, free-standing slab of absorbing material, whose
thickness and (complex) refractive index are properly tuned to implement the zero determinant
condition [\cite{WanScience2011}].
At the target resonance wavelength, light absorption in the slab can be modulated
in full by means of the input beams dephasing. For off-resonance wavelengths the modulation is
much smaller, and a CPA bandwidth can hence be defined.
To date, the majority of CPA devices that have been proposed exhibit a
strict permutation symmetry between the two input ports, and the proposals of wide-band CPA 
only have theoretical character. Moreover, the material loss mechanisms employed
feature a much larger bandwidth than the
energy exchange rate with the electromagnetic field.
Therefore, light-matter interaction can be described under a perturbative
approach, and the CPA bandwidth is connected
to the sole resonance decay time of the photonic resonator loaded by the non-radiative losses [\cite{ZanottoSperoNatPhys}].
On the contrary, when a material excitation (an atomic transition, for instance) is coupled to an optical resonator having a similar bandwidth and a sufficiently small modal volume, the strong coupling regime occurs [\cite{Kimble}]. In this situation new quantum eigenstates of the system are formed as a combination of material and photonic excitations (polaritons) and energy is continuously exchanged between the two “fields” at a rate corresponding to the coupling strength (the so-called vacuum Rabi energy) [\cite{KaluznyPRL1983,ThompsonPRL1992,WeisbuchPRL1992,DiniPRL2003,ReithmaierNat2004,YoshieNat2004,PeterPRL2005,AmoNat2009}]. One question now naturally arises: is it possible to feed perfectly and in full the electromagnetic energy from the outside world into such mixed light-matter polariton states? Or, in other words, is it possible to “critically couple” polaritons without destroying them? The question is intriguing in that the polariton concept intrinsically assumes a system that is nearly “closed” on the time scale of light-matter interaction, and one may intuitively expect an incompatibility of the strong-coupling conditions with the critical coupling ones.
The aim of this work is to review the recent efforts made to extend, both theoretically and experimentally, the concept of CPA to systems beyond the permutation symmetry hypothesis and the perturbative approach.
The next section is devoted to the theoretical treatment of two-port systems, assuming the preservation of reciprocity [\cite{Chew, DirkNatPhot2013}]; firstly an essential equation describing coherent absorption is derived, supported by a fast visual interpretation which reveals the main features of the prototypical two-port system. The second part gives a physical insight into the absorption mechanisms which lead a system to the CPA condition; to this aim a semiclassical approach is employed to develop a two-port coupled mode theory, where the matter degrees of freedom are coupled to the photonic ones. Physically sound parameters like the matter and radiative damping rates, the non radiative damping rate and the coupling rate allow to trace a phase diagram in their parametric space, where the emergence of two distinct CPA regimes is revealed.
In the third section the experimental strategies to reveal univocal signatures of the coherent absorption are illustrated, again divided in two complementary contributions, the first devoted to the general features of the two-port system, the second devoted to the features arising from the coupled mode theory (CMT). For each part, after the description of the typical devices and instrumental setups employed, the experimental results are presented and discussed, highlighting the unique characteristics related to coherent absorption and confronting them with the theoretical predictions.
A final section is dedicated to the concluding remarks and the future perspectives.

\section{Theoretical treatment of two-port systems}
\label{sec:1}
\subsection{Coherent Perfect Absorption in two-port systems}
\label{sec:2}
\begin{figure}[htbp]
\centering
\includegraphics[width=0.47\textwidth]{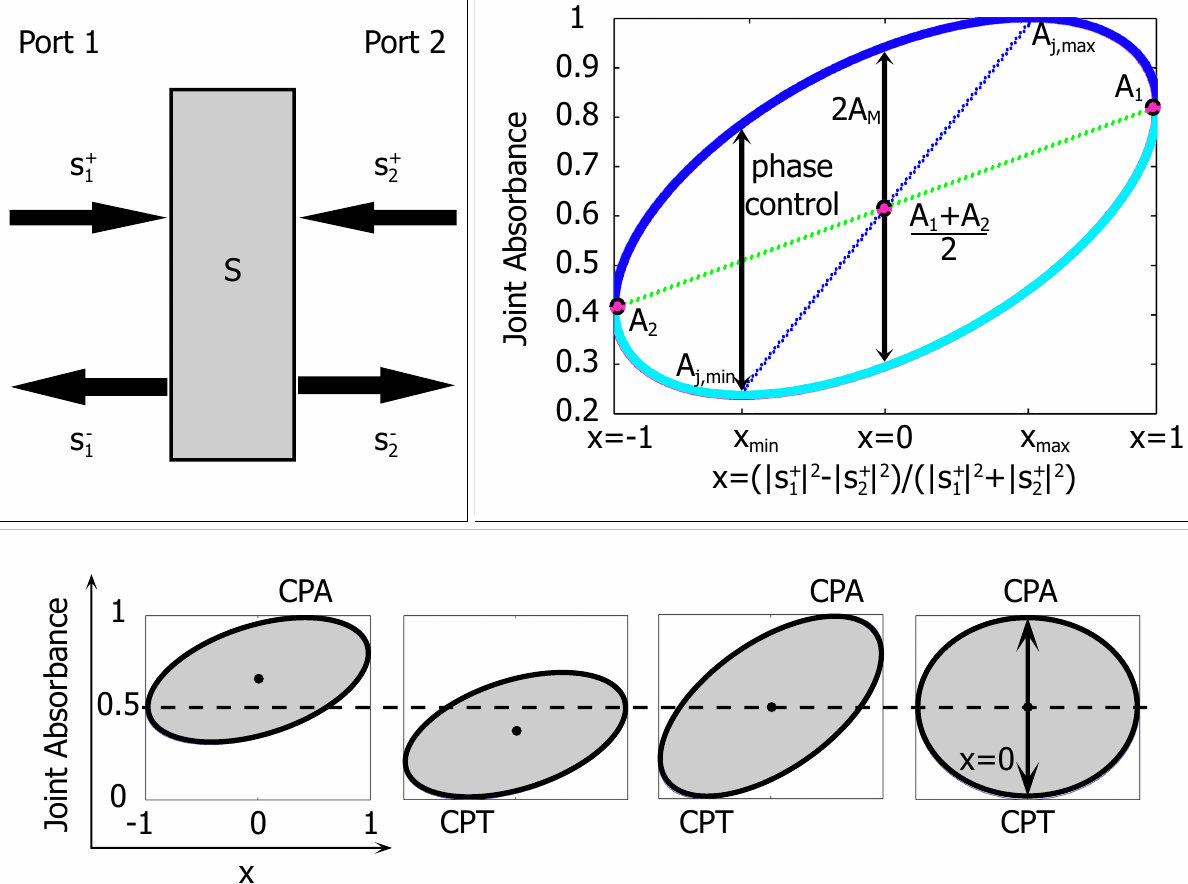}
\caption{Coherent absorption control in a linear two port scattering system with reciprocity. In the upper-left panel a general schematics is presented, with the scattering properties defined by a matrix S. If the two inputs are coherent, the total absorbance $A_j = 1-I_{out}/I_{in}$ can be controlled by means of their relative phase $\varphi = \arg{(s_2^+/s_1^+)}$ and access all values delimited by an ellipse as function of their normalised relative intensity $x = (\abs{s_1^+}^2 -\abs{s_2^+}^2) / (\abs{s_1^+}^2 + \abs{s_2^+}^2)$, as illustrated by the graph in the upper-right panel. For fixed $x$, absorbance can assume all different values along the vertical line delimited by the ellipse simply varying the phase difference $\varphi$. Instead, total absorbance in the presence of two incoherent inputs is described by a straightline connecting $A_2$ to $A_1$, with no possibility to be controlled by $\varphi$.}
\label{fig:SM_twoport}
\end{figure}

Consider a linear two-port system like that schematized in Fig.\ \ref{fig:SM_twoport}, upper left panel, driven by coherent input beams through the ports, i.e., through the scattering channels. Outside the device, the electromagnetic field can be described by the complex amplitudes of input and output fields, $s_{1,2}^{\pm}$. Provided that a proper normalization is employed, these amplitudes are connected with the electromagnetic power flux entering or leaving the device: $I_{1,2}^{\mathrm{in, out}} = \abs{s_{1,2}^{\pm}}^2$. Depending on the physical system, the input and output ports correspond to plane waves (in the case of a multilayer), to diffraction channels (in the case of a grating), to waveguide modes (in the case of integrated optical circuits), or to more complex far-field wave patterns (like the spatial harmonics in systems which exhibit a cylindrical or spherical symmetry). The theory that follows handles in a general way all the above situations, provided that only two input/output channels are coupled at a time.
In terms of the scattering matrix, input and output amplitudes are related by
\begin{equation}
\centering
\begin{pmatrix}
s_1^{-} \\
s_2^{-}
\end{pmatrix}
= S
\begin{pmatrix}
s_1^{+} \\
s_2^{+}
\end{pmatrix}
\label{eqn:SM1}
\end{equation}
where $S$ depends on the device structure and on the frequency of the monochromatic input beams.
Without loss of generality, if the system is reciprocal [\cite{Chew, DirkNatPhot2013}] the scattering matrix can be parametrized as
\begin{equation}
S =
e^{i\phi}
\begin{pmatrix}
\rho_1 e^{i\psi_1} & i\tau \\
i\tau & \rho_2 e^{i\psi_2} 
\end{pmatrix}
\label{eqn:SM2}
\end{equation}
where $\rho_{1,2}^2\equiv R_{1,2}$ are the reflectances and $\tau^2\equiv T$ is the transmittance of the device. Notice that for a system which is symmetric for the exchange of ports 1 and 2 one has $\rho_{1} = \rho_{2}$ and $\psi_1 = \psi_2$. 
Being interested in the energy fluxes through the system, it is useful to write down explicitly the total output intensity $I^{\mathrm{out}} = I^{\mathrm{out}}_1 + I^{\mathrm{out}}_2$, observed when the device is driven by two coherent fields dephased by $\varphi = \arg(s_{2}^{+}/s_{1}^{+})$:
\begin{equation}
\begin{aligned}
&I^{\mathrm{out}} = \abs{s_{1}^{-}}^2 + \abs{s_{2}^{-}}^2 =\\
&(\rho_1^2 + \tau^2)\abs{s_{1}^{+}}^2 + (\rho_2^2 + \tau^2)\abs{s_{2}^{+}}^2 + \\
&+ 2 \abs{s_{1}^{+}} \abs{s_{2}^{+}} \tau \sqrt{ \rho^2_1 + \rho^2_2 - 2\rho_1 \rho_2 \cos{(\psi_1 + \psi_2)} }\, \sin (\varphi + \delta).
\label{eqn:I_out}
\end{aligned}
\end{equation}
The last term accounts for the interference induced by the coherent nature of the input beams. It depends on the input field dephasing, but also on a device-specific phase $\delta$, which explicitly reads
\begin{equation}
\tan{\delta} = \frac{\rho_2\sin{(\psi_2)} + \rho_1\sin{(\psi_1)}}{\rho_2\cos{(\psi_2)} - \rho_1\cos{(\psi_1)}}.
\label{eqn:delta}
\end{equation}
Defining the \textit{joint absorbance} as the fraction of energy absorbed when the system is driven by the coherent beams, from Eq.\ \ref{eqn:I_out} and applying straightforward algebraic manipulation one obtains 
\begin{equation}
\begin{aligned}
&A_j = 1 - \frac{I^{\mathrm{out}}}{I^{\mathrm{in}}} = \\
&\frac{1 + x}{2}A_1 + \frac{1 - x}{2}A_2 - \sqrt{1 - x^2}\ A_M\sin{(\varphi+\delta)},
\end{aligned}
\label{eqn:A_conj_gamma}
\end{equation}
where 
\begin{equation}
A_M = \sqrt{(1 - A_1)(1 - A_2) - \abs*{\det{S}}^2}
\end{equation}
and 
\begin{equation}
x = \frac{|s_1^+|^2 - |s_2^+|^2}{|s_1^+|^2 + |s_2^+|^2}.
\label{eqn:var_x}
\end{equation}
Equation \ref{eqn:A_conj_gamma}, which describes an ellipse in the plane $(x,A_j)$ (see Fig.\ \ref{fig:SM_twoport}), encloses in a compact form a number of significant features. First, when the parameter $x$ equals $\pm 1$, i.e.\ when the system is driven from a single port, the joint absorbance reduces to the ordinary single-beam absorbances $A_{1,2} = 1 - \rho_{1,2}^2 - \tau^2$. As far as $\abs{x}$ is moved toward smaller values, i.e., when the excitation becomes more symmetric, the absorbance is modulated by the input beams dephasing; the depth of such modulation is quantified by the parameter $A_M$. This phenomenon can be referred to as \textit{coherent control of absorption}, and is a precursor of proper CPA. Analyzing the above equation, or Fig.\ \ref{fig:SM_twoport}, it turns out that the maximal absorption modulation always occurs when $x = 0$, regardless of the system asymmetry. Meanwhile, minimum and maximum joint absorbance are reached when $x = x_{\mathrm{min}, \mathrm{max}}$, a symmetrically placed pair of points whose explicit expression is given by
\begin{subequations}
\begin{equation*}
x_{\mathrm{min}} = - x_{\mathrm{max}}
\end{equation*}
\begin{equation*}
x_{\mathrm{max}} = \pm \sqrt{\frac{\left( A_1 - A_2 \right)^2}{4 A_M^2 + \left( A_1 - A_2 \right)^2}}
\end{equation*}
\label{eqn:max_coordinates}
\end{subequations}
where the plus (minus) sign applies when $A_1 > A_2$ ($A_1 < A_2$).
In general, the minimum and maximum values of joint absorbance are neither zero nor unity; indeed, it can be shown that unity absorbance occurs if and only if $\det{S} = 0$, i.e., when the usual condition for CPA holds [\cite{ChongPRL2010}] (Fig.\ \ref{fig:SM_twoport}, lower panel). In other words, when the conditions for CPA are met, a device which is only partially opaque under single-port operation can become fully absorbing under proper double-port driving\footnote{It should be noticed that a degenerate situation, in which $S$ is the zero matrix, can also occur: this case corresponds to two decoupled single-port perfect absorbers.}. The opposite situation is that in which a sample, opaque when excited from a single port, becomes transparent under coherent two-port excitation: this situation is referred to as \textit{coherent perfect transparency} (CPT), lower panel of Fig.\ \ref{fig:SM_twoport}. A device may also simultaneously show both CPA and CPT, but only a symmetric device allows for the sweep between CPA and CPT by a simple phase modulation. Recalling Eq.\ \ref{eqn:delta}, it can be noticed that such a symmetric device exhibits CPA when the input beam dephasing is either $\varphi = 0$ or $\varphi = \pi$, as known from previous reports in the literature [\cite{ChongPRL2010, WanScience2011}]. It is worth noting that the Eq. \eqref{eqn:A_conj_gamma} still holds for negative values of the single-beam absorbances, e.g. for the case of optical amplifiers or PT-symmetric devices [\cite{LonghiPRA2010, SunPRL14}].
By analyzing Figs.\ \ref{fig:SM_twoport} (lower panel), a set of constraints on the single-beam absorbances can also be deduced. Indeed, in order to achieve the CPA, a device which is not optically active must show $(A_1 + A_2)/2 \ge 0.5$; on the other hand, CPT is reachable only if $(A_1 + A_2)/2 \le 0.5$. The compresence of CPA and CPT requires the strict condition $(A_1 + A_2)/2 = 0.5$, which is a significant relation in that it connects quantities accessible in an ordinary single-port excitation experiment with the device response to coherent excitation. The above relations are only necessary but not sufficient in order to observe CPA and/or CPT, since proper phase relations on the scattering matrix coefficients must also be fulfilled. They can be employed however in fast pre-screening single-beam experiments. 
To make predictions on the scattering coefficients it is necessary to get into the absorption and scattering mechanisms of the systems, which will be discussed in the next part of this section.

\subsection{Strong critical coupling}
\label{sec:3}

\begin{figure*}[htbp]
\centering
\includegraphics[width=0.75\textwidth]{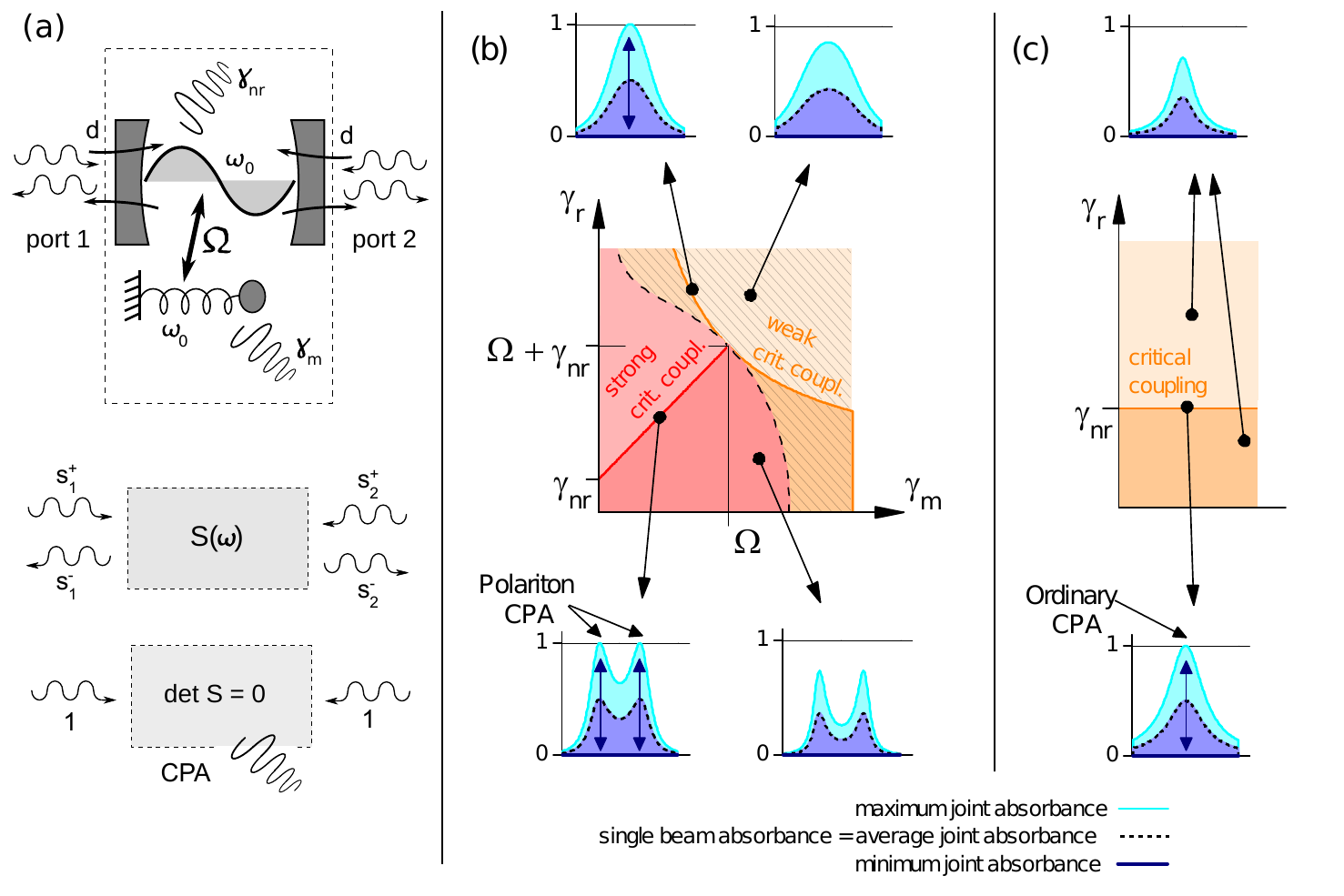}
\caption{Coupled oscillator model, strong/weak critical coupling, and polaritonic CPA. Panel (a): Two-port symmetric photonic resonator coupled to two scattering channels (ports) via the coupling constant $d$  (linked to the radiative damping rate $\gamma_r$), and to a matter resonator depicted as a spring-mass oscillator. Its linear response is given by the scattering matrix $S(\omega)$; when its determinant is zero all the incoming energy is absorbed (coherent perfect absorption, CPA). Panel (b), main graph: phase diagram of the coupled system. For small enough damping rates (flat-colored regions) the absorbance (subplots) exhibits two polaritonic peaks, while for large damping rates (striped regions) single-peaked spectra are observed. The dashed line marks the separation between the two regions. In both of these regions the joint absorbance can reach unity (CPA) provided that a critical coupling condition is fulfilled (see text). Panel (c): phase diagram with no cavity-matter coupling ($\Omega = 0$): the ordinary critical coupling condition $\gamma_r = \gamma_{nr}$ is recovered.}
\label{fig:Tredicucci_fig1}
\end{figure*}

A rigorous description of dissipation in a quantum picture would require a Green's function formalism [\cite{SavonaSSC1995}], or a master equation approach [\cite{AndreaniPRB1999,SrinivasanPRA2007,AuffevesPRA2007,ShenPRA2009_1,ShenPRA2009_2,DietzePRB2013}]. However, under the hypothesis of a small
average excitation density, a simpler semiclassical approach can be
employed. The matter degrees of freedom and the cavity photons
are represented as coupled harmonic oscillators, and a CMT yields equivalent analytical
formulas. This reveals that CPA, in either the weak- or strong-coupling
regime, is more deeply understood in terms of matching
of damping rates: no additional parameters need to be tailored.
In the following, the notation introduced
for photonic crystal resonances (Dirac notation) [\cite{FanJOSAA2003}] will be employed. The system
under study can now be sketched as in Fig. \ref{fig:Tredicucci_fig1}(a): it consists of an optical resonator
driven from the exterior via a coupling constant d, and the system
is assumed to be spatially symmetric. The optical cavity resonance
occurs at a pulsation $\omega_0$ and coherently exchanges energy with a
matter resonator oscillating at the same frequency via the coupling
constant. Energy is also re-radiated towards the exterior. The
amplitudes of the cavity and matter fields ($a$ and $b$ respectively), the
amplitudes of the incoming waves in the two ports $\ket{s^+} = (s_1^+,\,s_2^-)$,
and the corresponding outgoing wave amplitudes $\ket{s^-}$, are then
related by the following equations:
\begin{subequations}
\begin{equation}
\frac{db}{dt} = (i\omega_0 - \gamma_m)b + i\Omega a,
\end{equation}
\begin{equation}
\frac{da}{dt} = (i\omega_0 - \gamma_c)a + i\Omega b + (\bra{d}^*)\ket{s^+},
\end{equation}
\begin{equation}
\ket{s^-} = C\ket{s^+} + a\ket{d}.
\end{equation}
\end{subequations}
The evolution of the matter and cavity oscillators is damped by the presence of two decay channels, quantified respectively with the decay rates $\gamma_m$ and $\gamma_c$ . The latter is the cavity total damping rate, which is useful to separate into radiative and non-radiative contributions: $\gamma_c = \gamma_r + \gamma_{nr}$; $\gamma_r$ represents the radiative losses and $\gamma_{nr}$ keeps into account non-radiative, non-resonant cavity losses (for instance, free-carrier absorption and ohmic losses). 
In steady-state the system response is given by the frequency-dependent scattering matrix which connects the ingoing to the outgoing waves: $\ket{s^-} = S(\omega)\ket{s^+}$. Straightforward integration of the above equations yields:
\begin{equation}
S(\omega) = C - \frac{i(\omega - \omega_0) + \gamma_m}{(\omega - \omega_-)(\omega - \omega_+)}D.
\label{eqn:S_omega}
\end{equation}
The explicit expressions for matrices $C$ and $D=\ket{d}(\bra{d})^*$ can be found in Ref. [\cite{FanJOSAA2003}], while the polariton poles are 
\begin{equation*}
\omega_{\pm} = \omega_0 + \left[ i(\gamma_c + \gamma_m)\pm\sqrt{4\Omega^2 - (\gamma_c - \gamma_m)^2}\right]/2
\end{equation*}
Consistently with previous reports [\cite{ZanottoPRB2012}], this model predicts reflectance and transmittance lineshapes belonging to a Fano-like manifold; this complexity is lost, however, when dealing with the absorption properties, which are the object of the present work. It turns out then that a single spectral function is involved: $B(\omega) = (1 - \abs{\det{S(\omega)}}^2)$. The explicit expression of the S-matrix determinant following from \eqref{eqn:S_omega} is
\begin{equation*}
\det{S(\omega)} = \frac{(\omega - \bar{\omega}_+)(\omega - \bar{\omega}_-)}{(\omega - \omega_+)(\omega - \omega_-)},
\end{equation*}
with
\begin{equation*}
\begin{aligned}
\bar{\omega}_{\pm} = \omega_0 + \bigg[ i(-\gamma_r& + \gamma_{nr} + \gamma_m)+ \\
&\pm \sqrt{4\Omega^2 - (\gamma_r - \gamma_{nr} - \gamma_m)^2}\bigg]/2.
\end{aligned}
\end{equation*}
In a two-port system, the absorption is controlled by the simultaneous presence of two input beams; the relevant quantity is the joint absorbance $A_j$, defined as the ratio between absorbed and input energies when the two coherent beams excite the system ports with equal intensity. By sweeping the input beam dephasing $\varphi = \arg{(s_2^+ / s_1^+)}$, the joint absorbance sweeps from a minimum $A_{j,\, min}$ to a maximum $A_{j,\, max}$, which in the CMT model are given by
\begin{equation}
A_{j, \, min}(\omega) = 0; \quad A_{j,\, max}(\omega) = B(\omega).
\label{eqn:B_omega}
\end{equation}
Indeed, if $\det{S(\omega)} = 0$ one has $A_{j,\, max}(\omega) = 1$, i.e. a $\varphi$ can be found producing CPA. In addition, $A_{j,\, min}(\omega) = 0$ means that it always exists another $\varphi$ which gives coherent perfect transparency (CPT). In essence, a device implementing the two-oscillator CMT with a S-matrix satisfying $\det{S} = 0$ will behave as an ideal absorption-based interferometer.
If, on the contrary, a single beam excites the system either from port $1$ or $2$, the usual single-beam absorbances are observed. From the CMT it follows that in this case $A_1 (\omega) = A_2 (\omega) = B(\omega)/2$; this is consistent with the general theory of coherent absorption discussed previously, as it prescribes that the average between single-beam absorbances coincides with the average between minimum and maximum joint absorbances. This means, from one hand, that a sample whose single-beam absorption peaks at $1/2$ is expected to show CPA; on the other hand, single-beam absorbance never exceeds $1/2$ in a symmetric resonator supporting one resonant electromagnetic mode, as predicted by previous CMT calculations [\cite{ChutinanPRA2008}].  
The function $B(\omega)$ can exhibit either one or two peaks, depending on the region of the parameter space involved, as shown in Fig. \ref{fig:Tredicucci_fig1}(b); this transition can be regarded as a crossover between weak and strong coupling regimes. In both weak- and strong-coupling regions a curve in the parameter space exists corresponding to $\det{S(\omega)} = 0$, hence CPA. For small enough matter damping rate ($\gamma_m < \Omega$), one observes CPA when the damping rate matching 
\begin{equation}
\gamma_c = \gamma_{nr} + \gamma_m\,\,\textit{(strong critical coupling)}
\label{eqn:strong_critical_coupling}
\end{equation}
is fulfilled; notice the contribution of non-radiative losses which can help a good matter resonator entering the strong critical coupling [\cite{AuffevesPRB2010}]. On the other hand, the same system supports a second kind of critical coupling, which occurs when the following is satisfied:
\begin{equation}
\gamma_m(\gamma_c - \gamma_{nr}) = \Omega^2\,\, \textit{(weak critical coupling)}.
\end{equation}
In both weak and strong critical coupling all the incoming energy can be perfectly absorbed by the system, although it is obviously redistributed between the non-radiative cavity losses and the matter oscillator. A special case is found when the resonator-oscillator coupling $\Omega$ is set to zero: strong critical coupling no longer exists. Instead, the model shows that weak critical coupling reduces to the conventional CPA. This result is non-trivial, and it is also aesthetically gratifying: CPA is nothing else than critical coupling in a two-port system, and it obeys the usual critical coupling condition $\gamma_r = \gamma_{nr}$. This situation is shown in Fig. \ref{fig:Tredicucci_fig1}(c), where the absorption lineshapes given by $B(\omega)$ become purely Lorentzian. 

\section{Experimental design}
\label{sec:4}
\subsection{Experimental proof of the elliptical trace and broadband operation}
\begin{figure}[htbp]
\centering
\includegraphics[width=0.47\textwidth]{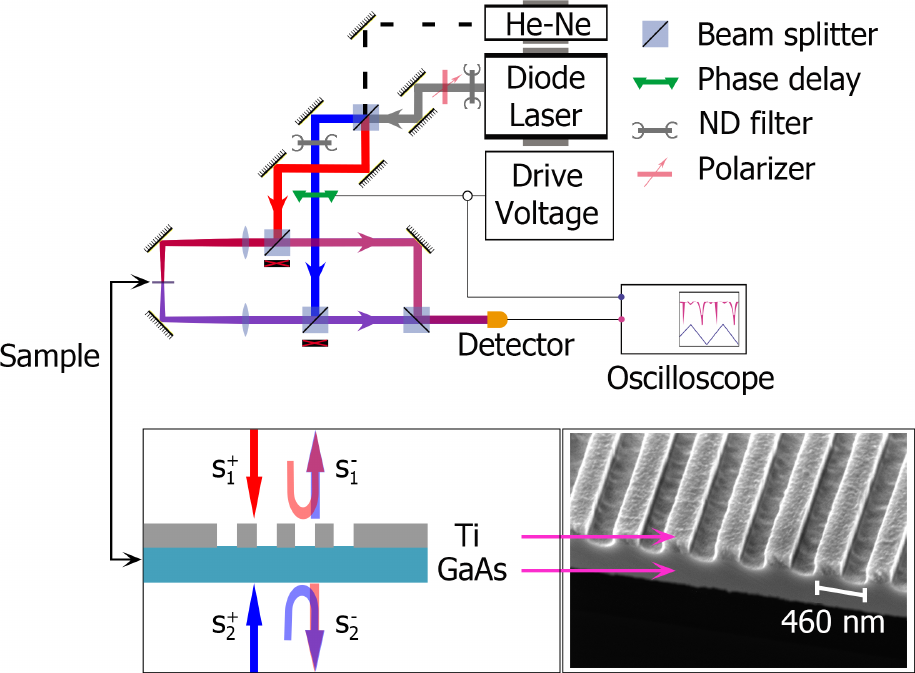}
\caption{Experimental setup and schematics of the sample. Radiation generated by a tunable diode laser is split into two and focused onto the opposite sides of the sample. The relative phase and intensity of the two input beams are controlled respectively by a neutral density filter wheel and a liquid crystal device driven by a voltage sweep. The total output intensity is then collected by a detector. A HeNe laser is employed for alignment purposes. Lower-right panel: SEM image of a cleaved sample.}
\label{fig:Exp_Setup}
\end{figure}

As a prototype of broadband coherent absorber, a thin plasmonic metamaterial was analized, based on the subwavelength grating geometry sketched in Fig.\ \ref{fig:Exp_Setup} . It consists of a periodic array of titanium stripes lying on a gallium arsenide suspended membrane. The device is characterized by titanium and gallium arsenide thicknesses $d_{\mathrm{Ti}}$ and $d_{\mathrm{GaAs}}$, by the periodicity $a$, and by the \textit{filling fraction} $f$, which is defined as the ratio between the volume fraction of titanium an the volume fraction of air.
If the device is driven by two counterpropagating beams normally incident on the surfaces, it can be considered as a two-port device in the sense that only the zeroth diffraction order propagates in the far field outside the sample, according to the Helmholtz equation in vacuum.
Inside, a complex interplay between modes supported by the metal-air and metal-semiconductor interfaces leads to a broadband absorption band centered in the 900 - 1000 nm wavelength range. The fine tuning of the sample parameters $d_{\mathrm{Ti}}$, $d_{\mathrm{GaAs}}$ and $a$ relies on a numerical optimization of the CPA-related parameters $A_M$ and $\det{S}$ introduced in the previous Section. As numerical modeling tool the rigorous coupled-wave analysis (RCWA) was employed, in an implementation which ensures fast convergence for the TM polarization, i.e., when it is imposed that the sole $H \neq 0$ field component is that parallel to the pattern stripes [\cite{GranetJOSAA1996}]. From the simulations it turned out that the optimal parameters are $d_{\mathrm{Ti}} = 169\ \mathrm{nm}$, $d_{\mathrm{GaAs}} = 366\ \mathrm{nm}$, and $a = 457\ \mathrm{nm}$. The filling fraction $f$ was not the object of optimization because $f = 0.5$ is the best value considering fabrication issues.
The sample is processed starting from a GaAs wafer, on which a $\simeq 500\ \mathrm{nm}$ thick $\mathrm{Al}_{0.5}\mathrm{Ga}_{0.5}\mathrm{As}$ layer, followed by the $366\ \mathrm{nm}$ $\mathrm{GaAs}$ layer, are grown by molecular beam epitaxy. Then, the Ti layer is deposited via thermal evaporation, and the periodic pattern is defined on a $200 \times 200\ \mu \mathrm{m}^2$ area by means of electron beam lithography and inductively-coupled plasma reactive ion etching. Finally, through a set of wet etching steps which exploit $\mathrm{Al}_{0.5}\mathrm{Ga}_{0.5}\mathrm{As}$ as an etch-stop layer, the membrane is locally freed from the substrate, and the plasmonic metamaterial can be optically accessed from both sides. 
Experimental proof of broadband coherent absorption modulation is provided by measuring the total intensity leaving the sample in presence of two counterpropagating input beams. To this aim a Mach Zehnder interferometer was built, as shown in Fig. \ref{fig:Exp_Setup}. As coherent light source, a Toptica DL100 grating-coupled diode laser, tunable from 908 to 983 nm, is employed; the beam is linearly polarized using a Glan Thompson $(200000:1)$ polarizer. After passing the first beamsplitter, the laser radiation is separated in two beams of approximately equal intensity. In order to explore the full dynamics predicted by the general theory of Section 2, the relative phase and intensity between the input beams has to be modulated. To this aim a liquid crystal phase delay modulator was placed across one of the two paths, and a filter wheel alternatively in one of the two paths. The two input beams are then focused using two achromatic doublets with $400\ \mathrm{mm}$ focal length, in a way that the two beam waists cover the sole patterned area of the sample. The output beams are eventually directed onto a detector; in order to avoid mutual interference of these beams on the detector surface, the two output paths are slightly displaced (not shown in the figure).
\begin{figure}
\centering
\includegraphics[width=0.4\textwidth]{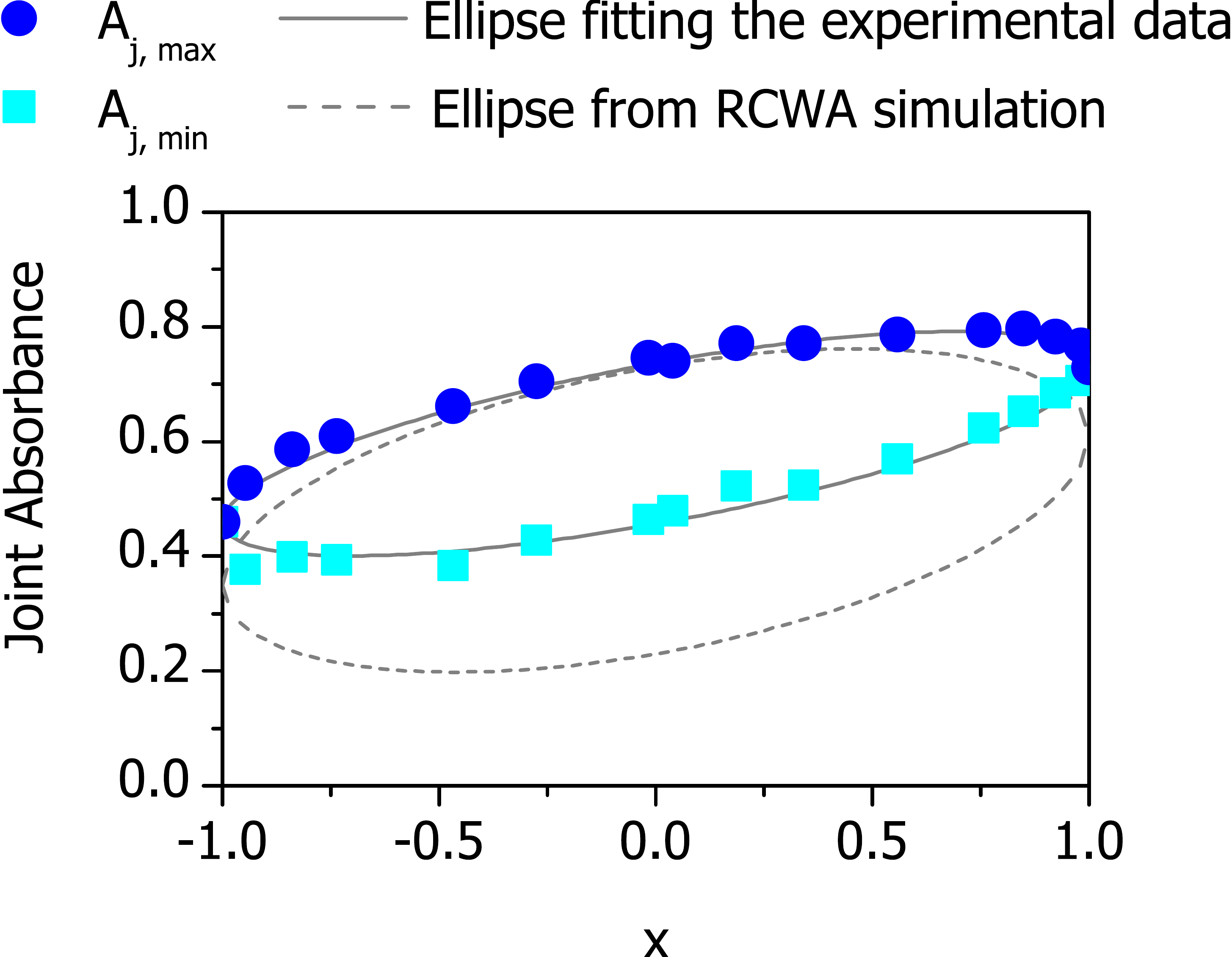}
\caption{Maximum and minimum joint absorbance as function of input beams relative intensity. The elliptical feature predicted in the general theory (Sect. 2 and Fig. \ \ref{fig:SM_twoport}) is clearly observed in the experiment. The dashed line is obtained through the rigorous coupled wave analysis (RCWA). The mismatch with the measured set is attributed to the roughness of the interfaces resulting from the etching processes, the bending of the sample due to the strain at the interface between Ti and GaAs, the complex spatial distribution of the incident beams, which RCWA does not account for, in addition to the possible mismatch between the dielectric constants employed, taken from the literature [\cite{JohnsonsChristy,Studna}], and their actual values.}
\label{fig:Exp_Ellipse}
\end{figure}
First proof of coherent control of absorption was observed for fixed laser wavelength $\lambda = 945\ \mathrm{nm}$. The results are shown in Fig. {\ref{fig:Exp_Ellipse}}, where the two beam relative intensity was plotted on the horizontal axis, and the joint absorbance on the vertical axis, in order to prove the theoretical predictions of Eq. \ref{eqn:A_conj_gamma}.
The circles and squares represent respectively the maximum and minimum experimental values of joint absorbance. Following Eq. \ref{eqn:A_conj_gamma}, two ellipses are calculated. The dashed gray line is the ellipse resulting when the absorption coefficients are calculated through RCWA analysis, while the solid gray line is obtained by using the experimental values for $A_1$, $A_2$ and $A_M$. The general agreement between the experimental features of the joint absorbance and the elliptical feature predicted in the general theory clearly appears (Sect. 2 and Fig. \ \ref{fig:SM_twoport}).
The disagreement between the RCWA calculations and the experimental results does not weaken the predictions of the general theory: the RCWA employed in this work derives the solution of Maxwell's equations for plane waves normally incident on a one dimensional grating made by perfectly vertical slabs, smooth surfaces, and materials having dielectric constants taken from the literature [\cite{JohnsonsChristy,Studna}]; more sophisticated methods which account for surface roughness and more complex spatial distribution of the incident radiation may improve the predictions on the fields and then on the scattering coefficients.
Nevertheless, even if the sample doesn't show the full features expected by design, the elliptical behavior of joint absorbance is preserved, as it is the signature of a general two port system, regardless of the particular values assumed for the scattering coefficients.
The solid line ellipse has a fairly different weight, as it is obtained from the experimental values of absorbance. Once the single-beam absorbances and the modulation depth are known the experimental ellipse can be completely described, giving a fast visual support in evaluating the setup alignment (and the accuracy of the measurement) by sweeping the relative intensity of the two incident beams.
\begin{figure}[htbp]
\centering
\includegraphics[width=0.47\textwidth]{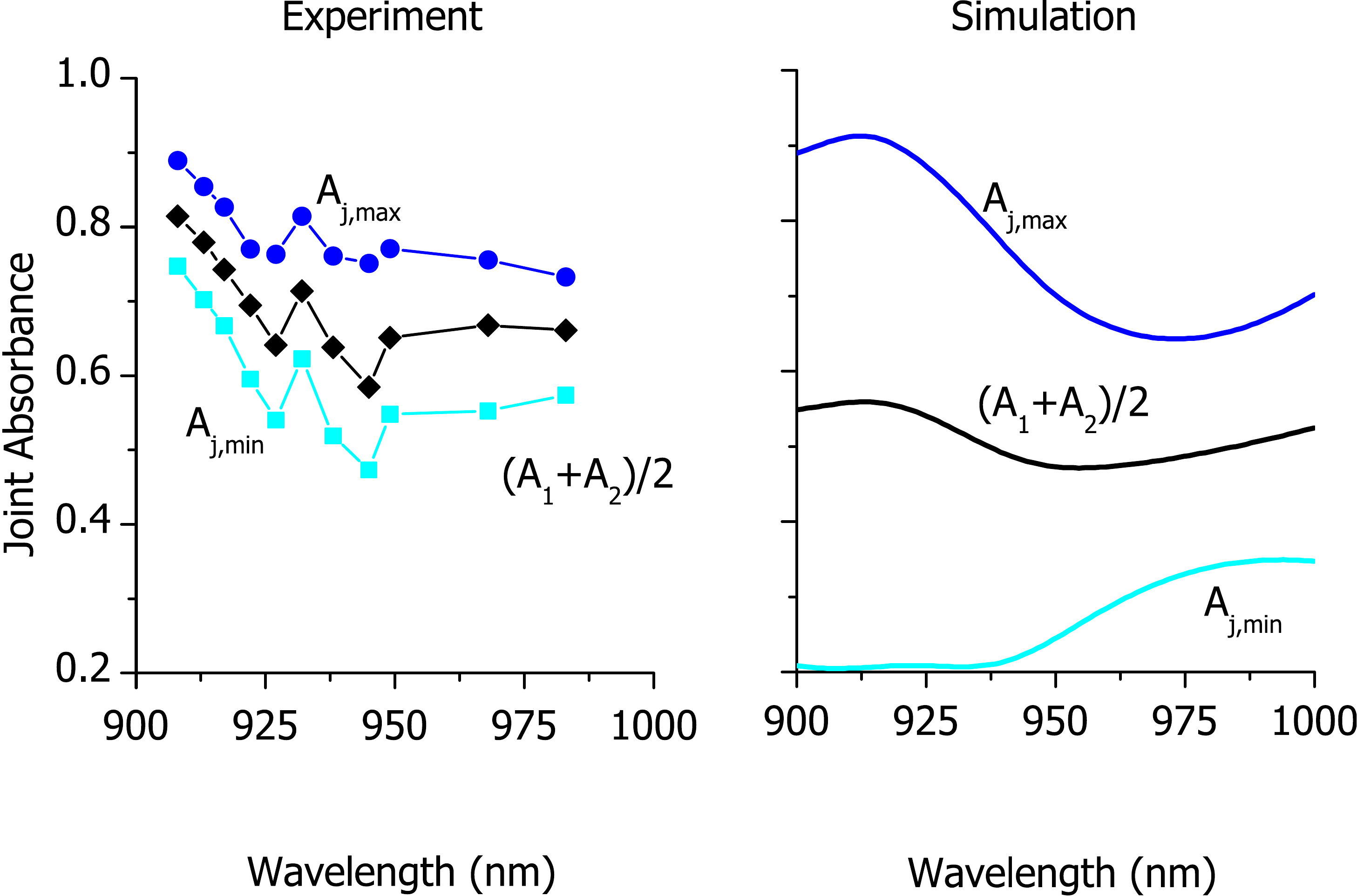}
\caption{Joint absorbance spectrum, experimental results (left) and theoretical predictions (right) obtained through RCWA. Incident beams have fixed equal intensity. In both cases the symmetry predicted in the general two port theory is preserved. The broad operation band is attributed to the plasmonic resonance of the sample.}
\label{fig:Exp_Joint}
\end{figure}

In the second part of the experiment the broadband operation of the plasmonic metamaterial is demonstrated; the two input beams intensities are fixed to be equal, while the wavelength is swept across the accessible spectral range of the laser source.
On the left side of Fig.\ \ref{fig:Exp_Joint}, the experimental data are reported; again the circles and squares represent the measured maximum/minimum absorbance values, while the diamonds show the measured values for $(A_1+A_2)/2$, i.e. the average of the single-beam absorbance values. This demonstrates that the symmetries reported in Fig. \ref{fig:Exp_Ellipse} are preserved for all the probed wavelengths. On the right side of Fig.\ \ref{fig:Exp_Joint}, the RCWA calculations are presented, which predict coherent absorption control through the whole broad spectrum thanks to the plasmonic nature of the e.m. resonance. A similar behaviour was observed experimentally for all the $11$ selected wavelengths. Given the wavelength with maximum modulation depth, the modulation bandwidth can be defined as the frequency interval whose extremes have the modulation halved.
Even if the modulation depth is not as pronounced as the RCWA predictions, a broad operation bandwidth is observed. The absorption is maximally modulated at $\lambda = 945\ \mathrm{nm}$, with an excursion of $30\%$, while the extremes of the spectrum, $\lambda = 908\ \mathrm{nm}$ and $\lambda = 983\ \mathrm{nm}$, still exhibit an excursion of $15\%$ and $20\%$ respectively. Thus the operation bandwidth of this device is about 25 THz, one hundred times the bandwidth reported for a silicon slab [\cite{WanScience2011}].
\begin{figure*}[htbp]
\centering
\includegraphics[width=0.75\textwidth]{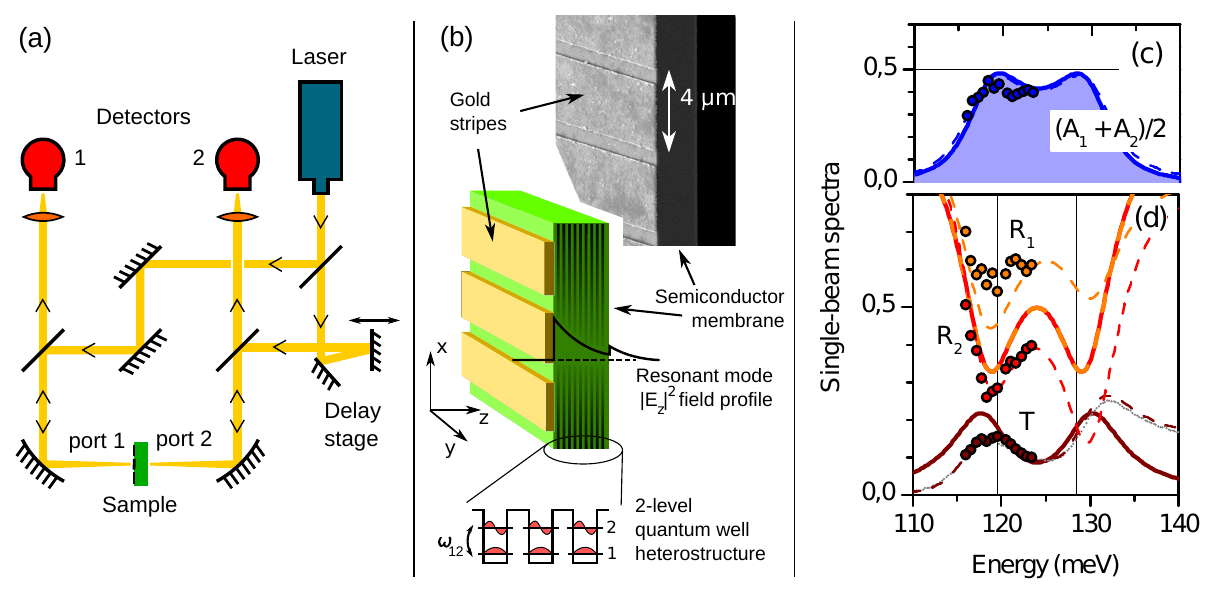}
\caption{Experimental setup, sample details and single-beam spectra of the polaritonic device. Panel (a): Interferometric arrangement for double-beam probing of the photonic crystal sample. The main features of the sample are outlined in panel (b): a thin semiconductor membrane is patterned with gold stripes implementing a metallic-dielectric photonic crystal resonator. In the membrane a multi-quantum well heterostructure acts as a collection of harmonic oscillators with resonance frequency $\omega_{12}$. Panels (c) and (d), large dots: experimental single-beam absorbance, reflectance and transmittance obtained by blocking one of the two interferometer’s input beams. Solid lines represent the CMT traces, while dashed lines are obtained via rigorous coupled-wave analysis (RCWA). The mismatch in reflectance and transmittance between CMT and experiment is possibly due to a second photonic resonance located at about $150 \, \mathrm{meV}$. This is also responsible for the difference between reflectances at ports $1$ and $2$. The transmittance was also measured with a broadband light source [small dots in panel (d)] to fully reveal the double-peak polaritonic structure; the perfect matching with the theoretical RCWA spectrum is here apparent.}
\label{fig:Tredicucci_fig2}
\end{figure*}

\subsection{Full interferometric control of absorption in polaritonic devices: strong critical coupling}
To experimentally demonstrate strong critical coupling and full interferometric control of absorption, an optical measurement  was performed on a polaritonic sample that mostly satisfies the radiative decay rate matching of Eq. \eqref{eqn:strong_critical_coupling}. The CPA setup is sketched in Fig. \ref{fig:Tredicucci_fig2}(a): the light source is a commercial external-cavity tunable quantum-cascade (QC) laser (Daylight Solutions) operating in the range of wavelengths $9.9  \,\mathrm{\mu m} \div 10.7  \,\mathrm{\mu m}$, the phase-delay stage is a loudspeaker membrane driven by a function generator, and the detectors are two liquid-nitrogen cooled mercury-cadmium-telluride devices. This is a consistent upgrade with respect to the previous setup (Fig. \ref{fig:Exp_Setup}), because the two detectors enable to detect separately the two contributions to the output intensity, $s_1^-$ and $s_2^-$. The sample consists of a semiconductor membrane which has been structured as a photonic crystal slab resonator (Fig. \ref{fig:Tredicucci_fig2}(b)); it embeds a multi-quantum well (MQW) heterostructure whose quantum design results in a single intersubband transition resonant with the photonic crystal mode, and hence in intersubband polariton states [\cite{DeglInnocentiSSC2011}]. In detail, a stack of 50 Al.33Ga.67As/GaAs QWs was employed, grown by molecular beam epitaxy, with barrier and well thicknesses of $30  \,\mathrm{nm}$ and $8.3  \,\mathrm{nm}$, respectively, and a nominal sheet doping $n = 5\cdot10^{11}  \,\mathrm{cm^{-2}}$ in the well material. The resulting $2  \,\mathrm{\mu m}$ thick membrane was patterned with metal stripes, featuring a period of $4  \,\mathrm{\mu m}$ and a duty cycle of $80 \%$, in order to tune the photonic resonance to the same energy of the intersubband transition and to satisfy the above condition \eqref{eqn:strong_critical_coupling}.  
Firstly, the sample has been characterized by means of single-beam spectroscopy, hence accessing the usual polaritonic spectra. The single-beam absorbance is plotted in Fig. \ref{fig:Tredicucci_fig2}(c) and shows that the first polaritonic peak (the only one accessible within the QC laser tuning range) is in good agreement with the theoretical curves obtained for the parameter set $\omega_0 = 124.5  \,\mathrm{meV}$, $\gamma_r = 3  \,\mathrm{meV}$, $\gamma_{nr} = 0$, $\gamma_m = 5  \,\mathrm{meV}$, $\Omega = 8 \,\mathrm{meV}$. The single-beam transmission was also probed with a broadband light source and Fourier transform interferometry [Fig. \ref{fig:Tredicucci_fig2}(d), small dots], in order to cover a wider spectral range revealing the characteristic polariton splitting and the perfect matching with theoretical modeling for both polariton peaks. The parameter values were also confirmed by independent RCWA calculations of the transmission by a photonic crystal bare resonator (transmission spectrum of a membrane with no active quantum wells) and by absorption measurements on the plain, unpatterned quantum well heterostructure. The proximity of the single-beam absorption peaks to the $50\%$ value is very promising for the observation of CPA; as explained above, in fact, in the framework of CMT this would automatically imply $\det{S(\omega)} = 0$ at those frequencies. As a matter of fact, the experimental points plotted in Fig. \ref{fig:Tredicucci_fig2}(c) are the average between the two single-beam absorbances, that slightly differ from each other in agreement with the differences in reflectance displayed in Fig. \ref{fig:Tredicucci_fig2}(d). This has to be attributed to the existence of a second resonant photonic mode, located at about $150\, \mathrm{meV}$, which is not included in the CMT model introduced here. On the contrary, this further resonance clearly appears in the RCWA transmission. This more accurate description then correctly reproduces the spectra of Fig. \ref{fig:Tredicucci_fig2}(d) (dashed lines); here, in fact, the structure is described ab initio starting from the geometrical parameters and the dielectric response of the semiconductor sample. The drawback of the RCWA model is that it does not provide any insight about the critical coupling in terms of damping rates matching. It has to be considered then as a complementary tool for a full interpretation of the experimental data. It is worth noting that the splitting between the polariton peaks in transmittance, reflectance and absorbance is not the same, as already pointed out in [\cite{SavonaSSC1995}].
\begin{figure}[htbp]
\centering
\includegraphics[width=0.47\textwidth]{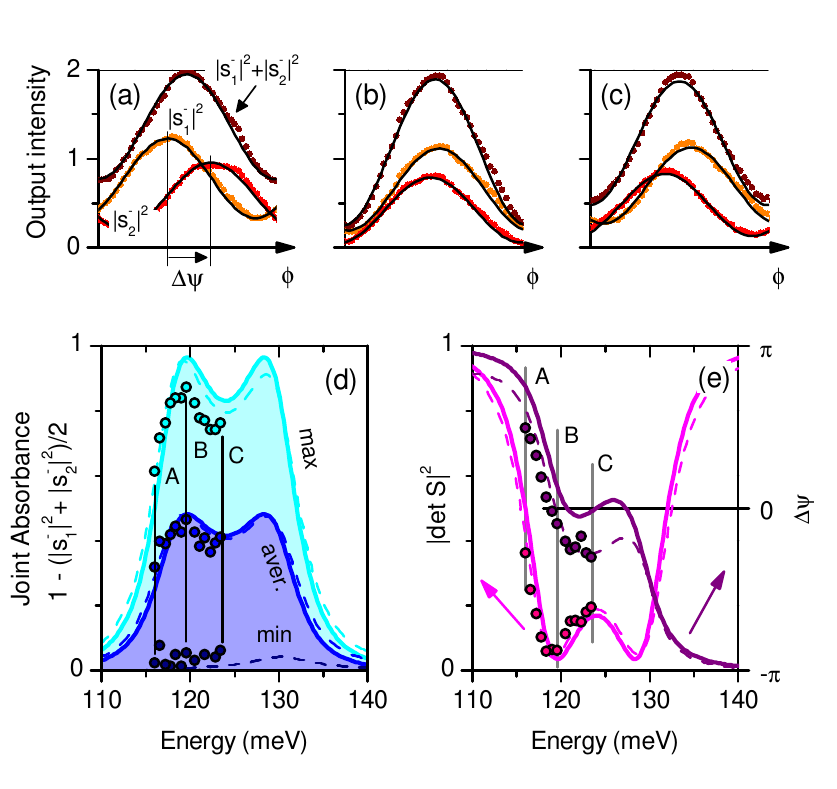}
\caption{Modulation of output intensity upon double-beam excitation, and polariton CPA. Panels (a-c): when the sample is excited from both ports $1$ and $2$ with unit-amplitude beams phase-shifted by $\varphi$, the output intensity recorded at ports $1$ and $2$ oscillates with a phase shift $\Delta\Psi$; the experimental traces (dots) are fitted by a sinusoid. Maximum, average, and minimum total output intensities are plotted in terms of joint absorbance [panel (d)], showing very good agreement with the two-peak polaritonic CPA theoretical traces (solid lines, CMT; dashed lines, rigorous coupled wave analysis). Further confirmation of the phenomenon is gained by analyzing the S-matrix determinant and the output beam dephasing [panel (e)]. Panels (a)-(c) correspond to the energies labeled A-C in panels (d) and (e).}
\label{fig:Tredicucci_fig3}
\end{figure}
The results of the double-beam experiment are reported in Fig. \ref{fig:Tredicucci_fig3}, panels (a) - (c), where the output intensities at ports $1$ and $2$ have been plotted, upon a sweep of the input beam phase difference $\varphi = \arg{(s2+/s1+)}$, for three different excitation wavelengths. In all three cases the total output intensity $\abs{s_1^-}^2 + \abs{s_2^-}^2$ reaches $2$, i.e. CPT (units are chosen such that input intensities $\abs{s_1^+}^2 = \abs{s_2^+}^2 = 1$); at the wavelength corresponding to (b) the $\varphi$-phase sweep also enables to reach $\abs{s_1^-}^2 + \abs{s_2^-}^2 \approx 0$, hence CPA. The corresponding joint absorbance values are reported in panel (d) as a function of the wavelength for the three phases $\varphi$ corresponding to minimum, average, and maximum output. The CMT and RCWA theoretical traces are also shown for comparison. It is worth noticing that the average joint absorbance, measured via the double-beam experiment, closely follows the average single-beam absorbances (Fig. \ref{fig:Tredicucci_fig2} (c)). The wavelength tuning of the laser covers basically the whole lower peak of the polariton doublet up to the resonance energy, unambiguously proving the polaritonic nature of the CPA and the existence of strong critical coupling. The occurrence of strong critical coupling in photonic crystal membrane intersubband polariton samples naturally follows from the very fact that in these samples the cavity and matter damping rates are easily close to each other. Indeed, it is a general feature following from the CMT that the modulation depth of the double-peaked absorption, and hence the strong critical coupling condition, is rather forgiving with respect to the relative damping rate mismatch. In this case, despite being $(\gamma_m - \gamma_r)/(\gamma_m + \gamma_r) \sim 25\%$, $A_{j, \,max} - A_{j,\, min} \approx 90\%$ was observed. Careful tailoring of $\gamma_r$ is anyway possible by tuning the grating parameters, for instance by a controlled etching of the slits in the underlying semiconductor [\cite{ManceauAPL2013}]. Further confirmation that the full coherent modulation of the absorption with phase difference is not an artifact due to other components in the setup is also given by the analysis of the output beam dephasing $\Delta\Psi$, which is connected to the phases of the scattering-matrix elements by the relation
\begin{equation*}
\Delta\Psi = \psi_1 + \psi_2 +\pi,
\end{equation*}
and hence represents a key feature of the sample under test. In panel (e) of Fig. \ref{fig:Tredicucci_fig3} the measured $\Delta\Psi$ is reported, and compared with the CMT and RCWA calculations. Excellent agreement with the latter is observed, while the difference with the CMT can be again attributed to the presence of the second photonic mode at higher energy.
In addition, an experimental estimate of the S-matrix determinant is provided, which does not rely on the link between $\det{S}$ and the absorbance (Eq. \eqref{eqn:B_omega}), valid only in this CMT model, but rather on the general relation
\begin{equation*}
\abs{\det{S}} = \left|T - e^{i\Delta\Psi}\sqrt{R_1 R_2}\right|.
\end{equation*}
Employing the single-beam reflectances  and transmittance, and the dephasing $\Delta\Psi$ extracted from the double-beam experiment, the curve reported in Fig. \ref{fig:Tredicucci_fig3}(e) is obtained, which shows excellent agreement with the theory.
\section{Conclusions and perspectives}
\label{sec:7}
In the CPA framework, an analysis on general two-port reciprocal systems has been developed: regardless of geometry, absorption features and far-field wavepattern, the joint absorption is represented by an ellipse determined by three experimentally accessible parameters: the absorptions related to single port input, and the modulation depth when the two input beams have equal intensity. This result gives an easy-to-use test for pre-screening and full characterization of samples whose working principle is the coherent control of absorption. To further investigate the physics of the internal absorption and scattering mechanisms,  a semiclassical model in terms of damping rates matching allowed to generalize the critical coupling condition - already known for lossy resonators - to the strong-coupling polariton framework, with potential applications to all the situations where two coupled oscillators are driven by two coherent sources.
Supported by the general two-port theory, broadband coherent absorption control was observed using a thin metallo-dielectric metamaterial, without the permutation symmetry between the two ports. Relying on its broad plasmonic resonances in the near infrared spectral region, an operational bandwidth of $\sim 1/10$ of the central wavelength was found.
Finally, the two-port CMT developed allowed optimal experimental design in order to observe coherent perfect absorption and coherent perfect transparency (and hence complete interferometric control of absorption) by dressed light-matter states.
The relevance of strong critical coupling rather than, for example, perfect transmission [\cite{ShenPRA2010}],  is motivated by
the need for efficient optical pumping in the proposal of
various intersubband polariton sources [\cite{DeLiberatoPRL2009,DeLiberatoPRB2013}]. Here, maximizing
the absorption into polariton states is of crucial importance, whilst
the polariton splitting needs to be precisely tailored to the optical
phonon resonance: the concept of CPA indeed enables one to
tune the absorption independently of the charge density, the main
parameter determining the splitting energy.
Further studies are necessary to explore the true quantum regime, in which pairs of individual photons are now used to drive the strongly coupled cavity-emitter system, which - ultimately - could feature a single quantum emitter. A generalization of the Hong-Ou-Mandel dip phenomenology has been predicted in the case of coherent perfect absorbers [\cite{BarnettPRA1998}], but this physics is still completely unexplored for polaritonic states.

\begin{acknowledgements}
This work was supported in part by the ERC advanced grant no. 321122 SouLMan
\end{acknowledgements}

% BibTeX users please use one of
%\bibliographystyle{spbasic}      % basic style, author-year citations
%\bibliographystyle{spmpsci}      % mathematics and physical sciences
%\bibliographystyle{spphys}       % APS-like style for physics
\bibliographystyle{aps-nameyear2}
\bibliography{ReviewLincei}   % name your BibTeX data base

\begin{thebibliography}{42}
% BibTex style file: aps.bst  (nameyear), 2013-04-23
\ifx \bisbn   \undefined \def \bisbn  #1{ISBN #1}\fi
\ifx \binits  \undefined \def \binits#1{#1} \fi
\ifx \bauthor  \undefined \def \bauthor#1{#1} \fi
\ifx \bjtitle  \undefined \def \bjtitle#1{\textrm{#1}}\fi
\ifx \batitle  \undefined \def \batitle#1{#1} \fi
\ifx \bctitle  \undefined \def \bctitle#1{#1} \fi
\ifx \bvolume  \undefined \def \bvolume#1{\textbf{#1}}\fi
\ifx \byear  \undefined \def \byear#1{#1} \fi
\ifx \bissue  \undefined \def \bissue#1{#1} \fi
\ifx \bfpage  \undefined \def \bfpage#1{#1} \fi
\ifx \blpage  \undefined \def \blpage #1{#1} \fi
\ifx \burl  \undefined \def \burl#1{#1} \fi
\ifx \doiurl  \undefined \def \doiurl#1{#1} \fi
\ifx \betal  \undefined \def \betal{et al.} \fi
\ifx \binstitute  \undefined \def \binstitute#1{#1} \fi
\ifx \beditor  \undefined \def \beditor#1{#1} \fi
\ifx \bpublisher  \undefined \def \bpublisher#1{#1} \fi
\ifx \bbtitle  \undefined \def \bbtitle#1{\textit{#1}} \fi
\ifx \bedition  \undefined \def \bedition#1{#1} \fi
\ifx \bseriesno  \undefined \def \bseriesno#1{#1} \fi
\ifx \blocation  \undefined \def \blocation#1{#1} \fi
\ifx \bsertitle  \undefined \def \bsertitle#1{#1} \fi
\ifx \bsnm \undefined \def \bsnm#1{#1} \fi
\ifx \bsuffix \undefined \def \bsuffix#1{#1} \fi
\ifx \bparticle \undefined \def \bparticle#1{#1} \fi
\ifx \barticle \undefined \def \barticle#1{#1} \fi
\ifx \botherref \undefined \def \botherref #1{#1} \fi
\ifx \url \undefined \def \url#1{#1} \fi
\ifx \bchapter \undefined \def \bchapter#1{#1} \fi
\ifx \bbook \undefined \def \bbook#1{#1} \fi
\ifx \bcomment \undefined \def \bcomment#1{#1} \fi
\ifx \oauthor \undefined \def \oauthor#1{#1} \fi
\ifx \citeauthoryear \undefined \def \citeauthoryear#1{#1} \fi
\ifx \texttildelow  \undefined \def \texttildelow{\symbol{126}} \fi
\def \endbibitem {}
\ifx \bconflocation  \undefined \def \bconflocation#1{#1} \fi

\bibitem[\protect\citeauthoryear{Amo et~al.}{2009}]{AmoNat2009}
\begin{barticle}
\bauthor{\binits{A.} \bsnm{Amo}},
\bauthor{\binits{D.} \bsnm{Sanvitto}},
\bauthor{\binits{F.} \bsnm{Laussy}},
\bauthor{\binits{D.} \bsnm{Ballarini}},
\bauthor{\binits{E.} \bsnm{Del~Valle}},
\bauthor{\binits{M.} \bsnm{Martin}},
\bauthor{\binits{A.} \bsnm{Lemaitre}},
\bauthor{\binits{J.} \bsnm{Bloch}},
\bauthor{\binits{D.} \bsnm{Krizhanovskii}},
\bauthor{\binits{M.} \bsnm{Skolnick}}, \betal,
\batitle{Collective fluid dynamics of a polariton condensate in a semiconductor
  microcavity}.
\bjtitle{Nature}
\bvolume{457}(\bissue{7227}),
\bfpage{291}--\blpage{295}
(\byear{2009})
\end{barticle}
\endbibitem

\bibitem[\protect\citeauthoryear{Andreani et~al.}{1999}]{AndreaniPRB1999}
\begin{barticle}
\bauthor{\binits{L.C.} \bsnm{Andreani}},
\bauthor{\binits{G.} \bsnm{Panzarini}},
\bauthor{\binits{J.-M.} \bsnm{G\'erard}},
\batitle{Strong-coupling regime for quantum boxes in pillar microcavities:
  Theory}.
\bjtitle{Phys. Rev. B}
\bvolume{60},
\bfpage{13276}--\blpage{13279}
(\byear{1999}).
doi:\doiurl{10.1103/PhysRevB.60.13276}
\end{barticle}
\endbibitem

\bibitem[\protect\citeauthoryear{Aspnes and Studna}{1983}]{Studna}
\begin{barticle}
\bauthor{\binits{D.} \bsnm{Aspnes}},
\bauthor{\binits{A.} \bsnm{Studna}},
\batitle{Dielectric functions and optical parameters of si, ge, gap, gaas,
  gasb, inp, inas, and insb from 1.5 to 6.0 ev}.
\bjtitle{Phys. Rev. B}
\bvolume{27}(\bissue{2}),
\bfpage{985}
(\byear{1983})
\end{barticle}
\endbibitem

\bibitem[\protect\citeauthoryear{Auff{\`e}ves et~al.}{2010}]{AuffevesPRB2010}
\begin{barticle}
\bauthor{\binits{A.} \bsnm{Auff{\`e}ves}},
\bauthor{\binits{D.} \bsnm{Gerace}},
\bauthor{\binits{J.-M.} \bsnm{G{\'e}rard}},
\bauthor{\binits{M.F.} \bsnm{Santos}},
\bauthor{\binits{L.} \bsnm{Andreani}},
\bauthor{\binits{J.-P.} \bsnm{Poizat}},
\batitle{Controlling the dynamics of a coupled atom-cavity system by pure
  dephasing}.
\bjtitle{Phys. Rev. B}
\bvolume{81}(\bissue{24}),
\bfpage{245419}
(\byear{2010})
\end{barticle}
\endbibitem

\bibitem[\protect\citeauthoryear{Auff\`eves-Garnier
  et~al.}{2007}]{AuffevesPRA2007}
\begin{barticle}
\bauthor{\binits{A.} \bsnm{Auff\`eves-Garnier}},
\bauthor{\binits{C.} \bsnm{Simon}},
\bauthor{\binits{J.-M.} \bsnm{G\'erard}},
\bauthor{\binits{J.-P.} \bsnm{Poizat}},
\batitle{Giant optical nonlinearity induced by a single two-level system
  interacting with a cavity in the purcell regime}.
\bjtitle{Phys. Rev. A}
\bvolume{75},
\bfpage{053823}
(\byear{2007}).
doi:\doiurl{10.1103/PhysRevA.75.053823}
\end{barticle}
\endbibitem

\bibitem[\protect\citeauthoryear{Barnett et~al.}{1998}]{BarnettPRA1998}
\begin{barticle}
\bauthor{\binits{S.M.} \bsnm{Barnett}},
\bauthor{\binits{J.} \bsnm{Jeffers}},
\bauthor{\binits{A.} \bsnm{Gatti}},
\bauthor{\binits{R.} \bsnm{Loudon}},
\batitle{Quantum optics of lossy beam splitters}.
\bjtitle{Phys. Rev. A}
\bvolume{57},
\bfpage{2134}--\blpage{2145}
(\byear{1998}).
doi:\doiurl{10.1103/PhysRevA.57.2134}
\end{barticle}
\endbibitem

\bibitem[\protect\citeauthoryear{Chew}{1995}]{Chew}
\begin{bbook}
\bauthor{\binits{W.C.} \bsnm{Chew}},
\bbtitle{Waves and fields in inhomogeneous media},
vol. \bseriesno{522}
(\bpublisher{IEEE press},
\blocation{NewYork}, \byear{1995})
\end{bbook}
\endbibitem

\bibitem[\protect\citeauthoryear{Chong et~al.}{2010}]{ChongPRL2010}
\begin{barticle}
\bauthor{\binits{Y.D.} \bsnm{Chong}},
\bauthor{\binits{L.} \bsnm{Ge}},
\bauthor{\binits{H.} \bsnm{Cao}},
\bauthor{\binits{A.D.} \bsnm{Stone}},
\batitle{Coherent perfect absorbers: Time-reversed lasers}.
\bjtitle{Phys. Rev. Lett.}
\bvolume{105},
\bfpage{053901}
(\byear{2010}).
doi:\doiurl{10.1103/PhysRevLett.105.053901}
\end{barticle}
\endbibitem

\bibitem[\protect\citeauthoryear{Chong and Stone}{2011}]{chong2011multiportCPA}
\begin{barticle}
\bauthor{\binits{Y.} \bsnm{Chong}},
\bauthor{\binits{A.D.} \bsnm{Stone}},
\batitle{Hidden black: Coherent enhancement of absorption in strongly
  scattering media}.
\bjtitle{Physical review letters}
\bvolume{107}(\bissue{16}),
\bfpage{163901}
(\byear{2011})
\end{barticle}
\endbibitem

\bibitem[\protect\citeauthoryear{Chutinan and John}{2008}]{ChutinanPRA2008}
\begin{barticle}
\bauthor{\binits{A.} \bsnm{Chutinan}},
\bauthor{\binits{S.} \bsnm{John}},
\batitle{Light trapping and absorption optimization in certain thin-film
  photonic crystal architectures}.
\bjtitle{Phys. Rev. A}
\bvolume{78},
\bfpage{023825}
(\byear{2008}).
doi:\doiurl{10.1103/PhysRevA.78.023825}
\end{barticle}
\endbibitem

\bibitem[\protect\citeauthoryear{De~Liberato and
  Ciuti}{2009}]{DeLiberatoPRL2009}
\begin{barticle}
\bauthor{\binits{S.} \bsnm{De~Liberato}},
\bauthor{\binits{C.} \bsnm{Ciuti}},
\batitle{Stimulated scattering and lasing of intersubband cavity polaritons}.
\bjtitle{Phys. Rev. Lett.}
\bvolume{102},
\bfpage{136403}
(\byear{2009}).
doi:\doiurl{10.1103/PhysRevLett.102.136403}
\end{barticle}
\endbibitem

\bibitem[\protect\citeauthoryear{De~Liberato et~al.}{2013}]{DeLiberatoPRB2013}
\begin{barticle}
\bauthor{\binits{S.} \bsnm{De~Liberato}},
\bauthor{\binits{C.} \bsnm{Ciuti}},
\bauthor{\binits{C.C.} \bsnm{Phillips}},
\batitle{Terahertz lasing from intersubband polariton-polariton scattering in
  asymmetric quantum wells}.
\bjtitle{Phys. Rev. B}
\bvolume{87},
\bfpage{241304}
(\byear{2013}).
doi:\doiurl{10.1103/PhysRevB.87.241304}
\end{barticle}
\endbibitem

\bibitem[\protect\citeauthoryear{Degl’Innocenti
  et~al.}{2011}]{DeglInnocentiSSC2011}
\begin{barticle}
\bauthor{\binits{R.} \bsnm{Degl’Innocenti}},
\bauthor{\binits{S.} \bsnm{Zanotto}},
\bauthor{\binits{a.} \bsnm{Tredicucci}},
\bauthor{\binits{G.} \bsnm{Biasiol}},
\bauthor{\binits{L.} \bsnm{Sorba}},
\batitle{{One-dimensional surface-plasmon gratings for the excitation of
  intersubband polaritons in suspended membranes}}.
\bjtitle{Solid State Communications}
\bvolume{151}(\bissue{23}),
\bfpage{1725}--\blpage{1727}
(\byear{2011}).
doi:\doiurl{10.1016/j.ssc.2011.09.002}
\end{barticle}
\endbibitem

\bibitem[\protect\citeauthoryear{Dietze et~al.}{2013}]{DietzePRB2013}
\begin{barticle}
\bauthor{\binits{D.} \bsnm{Dietze}},
\bauthor{\binits{K.} \bsnm{Unterrainer}},
\bauthor{\binits{J.} \bsnm{Darmo}},
\batitle{Role of geometry for strong coupling in active terahertz
  metamaterials}.
\bjtitle{Phys. Rev. B}
\bvolume{87}(\bissue{7}),
\bfpage{075324}
(\byear{2013})
\end{barticle}
\endbibitem

\bibitem[\protect\citeauthoryear{Dini et~al.}{2003}]{DiniPRL2003}
\begin{barticle}
\bauthor{\binits{D.} \bsnm{Dini}},
\bauthor{\binits{R.} \bsnm{K\"ohler}},
\bauthor{\binits{A.} \bsnm{Tredicucci}},
\bauthor{\binits{G.} \bsnm{Biasiol}},
\bauthor{\binits{L.} \bsnm{Sorba}},
\batitle{Microcavity polariton splitting of intersubband transitions}.
\bjtitle{Phys. Rev. Lett.}
\bvolume{90},
\bfpage{116401}
(\byear{2003}).
doi:\doiurl{10.1103/PhysRevLett.90.116401}
\end{barticle}
\endbibitem

\bibitem[\protect\citeauthoryear{Fan et~al.}{2003}]{FanJOSAA2003}
\begin{barticle}
\bauthor{\binits{S.} \bsnm{Fan}},
\bauthor{\binits{W.} \bsnm{Suh}},
\bauthor{\binits{J.D.} \bsnm{Joannopoulos}},
\batitle{Temporal coupled-mode theory for the fano resonance in optical
  resonators}.
\bjtitle{J. Opt. Soc. Am. A}
\bvolume{20}(\bissue{3}),
\bfpage{569}--\blpage{572}
(\byear{2003}).
doi:\doiurl{10.1364/JOSAA.20.000569}
\end{barticle}
\endbibitem

\bibitem[\protect\citeauthoryear{Ghebrebrhan et~al.}{2011}]{GhebrebrhanPRA2011}
\begin{barticle}
\bauthor{\binits{M.} \bsnm{Ghebrebrhan}},
\bauthor{\binits{P.} \bsnm{Bermel}},
\bauthor{\binits{Y.} \bsnm{Yeng}},
\bauthor{\binits{I.} \bsnm{Celanovic}},
\bauthor{\binits{M.} \bparticle{Solja\ifmmode \check{c}\else
  \v{c}\fi{}i\ifmmode~\acute{c}\else} \bsnm{\'{c}\fi{}}},
\bauthor{\binits{J.} \bsnm{Joannopoulos}},
\batitle{Tailoring thermal emission via $q$ matching of photonic crystal
  resonances}.
\bjtitle{Phys. Rev. A}
\bvolume{83},
\bfpage{033810}
(\byear{2011}).
doi:\doiurl{10.1103/PhysRevA.83.033810}
\end{barticle}
\endbibitem

\bibitem[\protect\citeauthoryear{Granet and Guizal}{1996}]{GranetJOSAA1996}
\begin{barticle}
\bauthor{\binits{G.} \bsnm{Granet}},
\bauthor{\binits{B.} \bsnm{Guizal}},
\batitle{Efficient implementation of the coupled-wave method for metallic
  lamellar gratings in tm polarization}.
\bjtitle{J. Opt. Soc. Am. A}
\bvolume{13}(\bissue{5}),
\bfpage{1019}--\blpage{1023}
(\byear{1996}).
doi:\doiurl{10.1364/JOSAA.13.001019}
\end{barticle}
\endbibitem

\bibitem[\protect\citeauthoryear{Haus}{1984}]{Haus}
\begin{bbook}
\bauthor{\binits{H.A.} \bsnm{Haus}},
\bbtitle{Waves and fields in optoelectronics}
(\bpublisher{Prentice-Hall},
\blocation{Englewood Cliffs, NJ}, \byear{1984})
\end{bbook}
\endbibitem

\bibitem[\protect\citeauthoryear{Jalas et~al.}{2013}]{DirkNatPhot2013}
\begin{barticle}
\bauthor{\binits{D.} \bsnm{Jalas}},
\bauthor{\binits{A.} \bsnm{Petrov}},
\bauthor{\binits{M.} \bsnm{Eich}},
\bauthor{\binits{W.} \bsnm{Freude}},
\bauthor{\binits{S.} \bsnm{Fan}},
\bauthor{\binits{Z.} \bsnm{Yu}},
\bauthor{\binits{R.} \bsnm{Baets}},
\bauthor{\binits{M.} \bsnm{Popovic}},
\bauthor{\binits{A.} \bsnm{Melloni}},
\bauthor{\binits{J.D.} \bsnm{Joannopoulos}},
\bauthor{\binits{M.} \bsnm{Vanwolleghem}},
\bauthor{\binits{C.R.} \bsnm{Doerr}},
\bauthor{\binits{H.} \bsnm{Renner}},
\batitle{What is -- and what is not -- an optical isolator}.
\bjtitle{Nature Photonics}
\bvolume{7}(\bissue{8}),
\bfpage{579}--\blpage{582}
(\byear{2013}).
doi:\doiurl{http://dx.doi.org/10.1038/nphoton.2013.185}
\end{barticle}
\endbibitem

\bibitem[\protect\citeauthoryear{Johnson and Christy}{1972}]{JohnsonsChristy}
\begin{barticle}
\bauthor{\binits{P.B.} \bsnm{Johnson}},
\bauthor{\binits{R.-W.} \bsnm{Christy}},
\batitle{Optical constants of the noble metals}.
\bjtitle{Phys. Rev. B}
\bvolume{6}(\bissue{12}),
\bfpage{4370}
(\byear{1972})
\end{barticle}
\endbibitem

\bibitem[\protect\citeauthoryear{Kaluzny et~al.}{1983}]{KaluznyPRL1983}
\begin{barticle}
\bauthor{\binits{Y.} \bsnm{Kaluzny}},
\bauthor{\binits{P.} \bsnm{Goy}},
\bauthor{\binits{M.} \bsnm{Gross}},
\bauthor{\binits{J.M.} \bsnm{Raimond}},
\bauthor{\binits{S.} \bsnm{Haroche}},
\batitle{Observation of self-induced rabi oscillations in two-level atoms
  excited inside a resonant cavity: The ringing regime of superradiance}.
\bjtitle{Phys. Rev. Lett.}
\bvolume{51},
\bfpage{1175}--\blpage{1178}
(\byear{1983}).
doi:\doiurl{10.1103/PhysRevLett.51.1175}
\end{barticle}
\endbibitem

\bibitem[\protect\citeauthoryear{Kimble}{1994}]{Kimble}
\begin{bchapter}
\bauthor{\binits{H.J.} \bsnm{Kimble}},
\bctitle{Structure and Dynamics in Cavity Quantum Electrodynamics},
in \bbtitle{Cavity Quantum Electrodynamics},
ed. by \beditor{\binits{P.R.} \bsnm{Berman}}
(\bpublisher{Academic Press},
\blocation{Boston}, \byear{1994})
\end{bchapter}
\endbibitem

\bibitem[\protect\citeauthoryear{Longhi}{2010}]{LonghiPRA2010}
\begin{barticle}
\bauthor{\binits{S.} \bsnm{Longhi}},
\batitle{Pt-symmetric laser absorber}.
\bjtitle{Phys. Rev. A}
\bvolume{82},
\bfpage{031801}
(\byear{2010}).
doi:\doiurl{10.1103/PhysRevA.82.031801}
\end{barticle}
\endbibitem

\bibitem[\protect\citeauthoryear{Manceau et~al.}{2013}]{ManceauAPL2013}
\begin{barticle}
\bauthor{\binits{J.-M.} \bsnm{Manceau}},
\bauthor{\binits{S.} \bsnm{Zanotto}},
\bauthor{\binits{I.} \bsnm{Sagnes}},
\bauthor{\binits{G.} \bsnm{Beaudoin}},
\bauthor{\binits{R.} \bsnm{Colombelli}},
\batitle{Optical critical coupling into highly confining metal-insulator-metal
  resonators}.
\bjtitle{Appl. Phys. Lett.}
\bvolume{103}(\bissue{9}),
\bfpage{091110}
(\byear{2013})
\end{barticle}
\endbibitem

\bibitem[\protect\citeauthoryear{Noh et~al.}{2012}]{noh2012CPA}
\begin{barticle}
\bauthor{\binits{H.} \bsnm{Noh}},
\bauthor{\binits{Y.} \bsnm{Chong}},
\bauthor{\binits{A.D.} \bsnm{Stone}},
\bauthor{\binits{H.} \bsnm{Cao}},
\batitle{Perfect coupling of light to surface plasmons by coherent absorption}.
\bjtitle{Phys. Rev. Lett.}
\bvolume{108}(\bissue{18}),
\bfpage{186805}
(\byear{2012})
\end{barticle}
\endbibitem

\bibitem[\protect\citeauthoryear{Peter et~al.}{2005}]{PeterPRL2005}
\begin{barticle}
\bauthor{\binits{E.} \bsnm{Peter}},
\bauthor{\binits{P.} \bsnm{Senellart}},
\bauthor{\binits{D.} \bsnm{Martrou}},
\bauthor{\binits{A.} \bsnm{Lemaitre}},
\bauthor{\binits{J.} \bsnm{Hours}},
\bauthor{\binits{J.} \bsnm{Gerard}},
\bauthor{\binits{J.} \bsnm{Bloch}},
\batitle{Exciton-photon strong-coupling regime for a single quantum dot
  embedded in a microcavity}.
\bjtitle{Phys. Rev. Lett.}
\bvolume{95},
\bfpage{067401}
(\byear{2005}).
doi:\doiurl{10.1103/PhysRevLett.95.067401}
\end{barticle}
\endbibitem

\bibitem[\protect\citeauthoryear{Reithmaier et~al.}{2004}]{ReithmaierNat2004}
\begin{barticle}
\bauthor{\binits{J.P.} \bsnm{Reithmaier}},
\bauthor{\binits{G.} \bsnm{Sek}},
\bauthor{\binits{A.} \bsnm{Loffler}},
\bauthor{\binits{C.} \bsnm{Hofmann}},
\bauthor{\binits{S.} \bsnm{Kuhn}},
\bauthor{\binits{S.} \bsnm{Reitzenstein}},
\bauthor{\binits{L.V.} \bsnm{Keldysh}},
\bauthor{\binits{V.D.} \bsnm{Kulakovskii}},
\bauthor{\binits{T.L.} \bsnm{Reinecke}},
\bauthor{\binits{A.} \bsnm{Forchel}},
\batitle{Strong coupling in a single quantum dot-semiconductor microcavity
  system}.
\bjtitle{Nature}
\bvolume{432},
\bfpage{197}
(\byear{2004}).
doi:\doiurl{10.1038/nature02969}
\end{barticle}
\endbibitem

\bibitem[\protect\citeauthoryear{Savona et~al.}{1995}]{SavonaSSC1995}
\begin{barticle}
\bauthor{\binits{V.} \bsnm{Savona}},
\bauthor{\binits{L.C.} \bsnm{Andreani}},
\bauthor{\binits{P.} \bsnm{Schwendimann}},
\bauthor{\binits{A.} \bsnm{Quattropani}},
\batitle{Quantum well excitons in semiconductor microcavities: Unified
  treatment of weak and strong coupling regimes}.
\bjtitle{Solid State Communications}
\bvolume{93}(\bissue{9}),
\bfpage{733}--\blpage{739}
(\byear{1995}).
doi:\doiurl{http://dx.doi.org/10.1016/0038-1098(94)00865-5}
\end{barticle}
\endbibitem

\bibitem[\protect\citeauthoryear{Shen and Fan}{2009a}]{ShenPRA2009_1}
\begin{barticle}
\bauthor{\binits{J.-T.} \bsnm{Shen}},
\bauthor{\binits{S.} \bsnm{Fan}},
\batitle{Theory of single-photon transport in a single-mode waveguide. i.
  coupling to a cavity containing a two-level atom}.
\bjtitle{Phys. Rev. A}
\bvolume{79},
\bfpage{023837}
(\byear{2009}a).
doi:\doiurl{10.1103/PhysRevA.79.023837}
\end{barticle}
\endbibitem

\bibitem[\protect\citeauthoryear{Shen and Fan}{2009b}]{ShenPRA2009_2}
\begin{barticle}
\bauthor{\binits{J.-T.} \bsnm{Shen}},
\bauthor{\binits{S.} \bsnm{Fan}},
\batitle{Theory of single-photon transport in a single-mode waveguide. ii.
  coupling to a whispering-gallery resonator containing a two-level atom}.
\bjtitle{Phys. Rev. A}
\bvolume{79},
\bfpage{023838}
(\byear{2009}b).
doi:\doiurl{10.1103/PhysRevA.79.023838}
\end{barticle}
\endbibitem

\bibitem[\protect\citeauthoryear{Shen and Fan}{2010}]{ShenPRA2010}
\begin{barticle}
\bauthor{\binits{J.-T.} \bsnm{Shen}},
\bauthor{\binits{S.} \bsnm{Fan}},
\batitle{Quantum critical coupling conditions for zero single-photon
  transmission through a coupled atom-resonator-waveguide system}.
\bjtitle{Phys. Rev. A}
\bvolume{82},
\bfpage{021802}
(\byear{2010}).
doi:\doiurl{10.1103/PhysRevA.82.021802}
\end{barticle}
\endbibitem

\bibitem[\protect\citeauthoryear{Srinivasan and
  Painter}{2007}]{SrinivasanPRA2007}
\begin{barticle}
\bauthor{\binits{K.} \bsnm{Srinivasan}},
\bauthor{\binits{O.} \bsnm{Painter}},
\batitle{Mode coupling and cavity\char21{}quantum-dot interactions in a
  fiber-coupled microdisk cavity}.
\bjtitle{Phys. Rev. A}
\bvolume{75},
\bfpage{023814}
(\byear{2007}).
doi:\doiurl{10.1103/PhysRevA.75.023814}
\end{barticle}
\endbibitem

\bibitem[\protect\citeauthoryear{Sun et~al.}{2014}]{SunPRL14}
\begin{barticle}
\bauthor{\binits{Y.} \bsnm{Sun}},
\bauthor{\binits{W.} \bsnm{Tan}},
\bauthor{\binits{H.-q.} \bsnm{Li}},
\bauthor{\binits{J.} \bsnm{Li}},
\bauthor{\binits{H.} \bsnm{Chen}},
\batitle{Experimental demonstration of a coherent perfect absorber with pt
  phase transition}.
\bjtitle{Phys. Rev. Lett.}
\bvolume{112},
\bfpage{143903}
(\byear{2014}).
doi:\doiurl{10.1103/PhysRevLett.112.143903}
\end{barticle}
\endbibitem

\bibitem[\protect\citeauthoryear{Thompson et~al.}{1992}]{ThompsonPRL1992}
\begin{barticle}
\bauthor{\binits{R.J.} \bsnm{Thompson}},
\bauthor{\binits{G.} \bsnm{Rempe}},
\bauthor{\binits{H.J.} \bsnm{Kimble}},
\batitle{Observation of normal-mode splitting for an atom in an optical
  cavity}.
\bjtitle{Phys. Rev. Lett.}
\bvolume{68},
\bfpage{1132}--\blpage{1135}
(\byear{1992}).
doi:\doiurl{10.1103/PhysRevLett.68.1132}
\end{barticle}
\endbibitem

\bibitem[\protect\citeauthoryear{Wan et~al.}{2011}]{WanScience2011}
\begin{barticle}
\bauthor{\binits{W.} \bsnm{Wan}},
\bauthor{\binits{Y.} \bsnm{Chong}},
\bauthor{\binits{L.} \bsnm{Ge}},
\bauthor{\binits{H.} \bsnm{Noh}},
\bauthor{\binits{A.D.} \bsnm{Stone}},
\bauthor{\binits{H.} \bsnm{Cao}},
\batitle{Time-reversed lasing and interferometric control of absorption}.
\bjtitle{Science}
\bvolume{331}(\bissue{6019}),
\bfpage{889}--\blpage{892}
(\byear{2011}).
doi:\doiurl{10.1126/science.1200735}
\end{barticle}
\endbibitem

\bibitem[\protect\citeauthoryear{Weisbuch et~al.}{1992}]{WeisbuchPRL1992}
\begin{barticle}
\bauthor{\binits{C.} \bsnm{Weisbuch}},
\bauthor{\binits{M.} \bsnm{Nishioka}},
\bauthor{\binits{A.} \bsnm{Ishikawa}},
\bauthor{\binits{Y.} \bsnm{Arakawa}},
\batitle{Observation of the coupled exciton-photon mode splitting in a
  semiconductor quantum microcavity}.
\bjtitle{Phys. Rev. Lett.}
\bvolume{69},
\bfpage{3314}--\blpage{3317}
(\byear{1992}).
doi:\doiurl{10.1103/PhysRevLett.69.3314}
\end{barticle}
\endbibitem

\bibitem[\protect\citeauthoryear{Yariv and Yeh}{2007}]{Yariv}
\begin{bbook}
\bauthor{\binits{A.} \bsnm{Yariv}},
\bauthor{\binits{P.} \bsnm{Yeh}},
\bbtitle{Photonics: Optical Electronics in Modern Communication}
(\bpublisher{Oxford University Press},
\blocation{New York}, \byear{2007})
\end{bbook}
\endbibitem

\bibitem[\protect\citeauthoryear{Yoshie et~al.}{2004}]{YoshieNat2004}
\begin{barticle}
\bauthor{\binits{T.} \bsnm{Yoshie}},
\bauthor{\binits{A.} \bsnm{Scherer}},
\bauthor{\binits{J.} \bsnm{Hendrickson}},
\bauthor{\binits{G.} \bsnm{Khitrova}},
\bauthor{\binits{H.M.} \bsnm{Gibbs}},
\bauthor{\binits{G.} \bsnm{Rupper}},
\bauthor{\binits{C.} \bsnm{Ell}},
\bauthor{\binits{O.B.} \bsnm{Shchekin}},
\bauthor{\binits{D.G.} \bsnm{Deppe}},
\batitle{Vacuum rabi splitting with a single quantum dot in a photonic crystal
  nanocavity}.
\bjtitle{Nature}
\bvolume{432},
\bfpage{200}
(\byear{2004}).
doi:\doiurl{10.1038/nature03119}
\end{barticle}
\endbibitem

\bibitem[\protect\citeauthoryear{Yu et~al.}{2012}]{YuPRL2012}
\begin{barticle}
\bauthor{\binits{Z.} \bsnm{Yu}},
\bauthor{\binits{A.} \bsnm{Raman}},
\bauthor{\binits{S.} \bsnm{Fan}},
\batitle{Thermodynamic upper bound on broadband light coupling with photonic
  structures}.
\bjtitle{Phys. Rev. Lett.}
\bvolume{109},
\bfpage{173901}
(\byear{2012}).
doi:\doiurl{10.1103/PhysRevLett.109.173901}
\end{barticle}
\endbibitem

\bibitem[\protect\citeauthoryear{Zanotto et~al.}{2012}]{ZanottoPRB2012}
\begin{barticle}
\bauthor{\binits{S.} \bsnm{Zanotto}},
\bauthor{\binits{R.} \bsnm{Degl'Innocenti}},
\bauthor{\binits{L.} \bsnm{Sorba}},
\bauthor{\binits{A.} \bsnm{Tredicucci}},
\bauthor{\binits{G.} \bsnm{Biasiol}},
\batitle{Analysis of line shapes and strong coupling with intersubband
  transitions in one-dimensional metallodielectric photonic crystal slabs}.
\bjtitle{Phys. Rev. B}
\bvolume{85},
\bfpage{035307}
(\byear{2012}).
doi:\doiurl{10.1103/PhysRevB.85.035307}
\end{barticle}
\endbibitem

\bibitem[\protect\citeauthoryear{Zanotto et~al.}{2014}]{ZanottoSperoNatPhys}
\begin{barticle}
\bauthor{\binits{S.} \bsnm{Zanotto}},
\bauthor{\binits{F.P.} \bsnm{Mezzapesa}},
\bauthor{\binits{F.} \bsnm{Bianco}},
\bauthor{\binits{G.} \bsnm{Biasiol}},
\bauthor{\binits{L.} \bsnm{Baldacci}},
\bauthor{\binits{M.S.} \bsnm{Vitiello}},
\bauthor{\binits{L.} \bsnm{Sorba}},
\bauthor{\binits{R.} \bsnm{Colombelli}},
\bauthor{\binits{A.} \bsnm{Tredicucci}},
\batitle{Perfect energy-feeding into strongly coupled systems and
  interferometric control of polariton absorption}.
\bjtitle{Nature Phys.}
\bvolume{10}(\bissue{11}),
\bfpage{830}--\blpage{834}
(\byear{2014}).
doi:\doiurl{10.1038/NPHYS3106}
\end{barticle}
\endbibitem

\end{thebibliography}
\end{document}